\documentclass[12pt]{iopart}

\usepackage{iopams}
 \usepackage{graphicx}
\usepackage{cite}

\usepackage[latin1]{inputenc}
\newcommand{\be}{\begin{equation}}
\newcommand{\ee}{\end{equation}}
\newcommand{\eqref}{\eref}

\def\LM#1#2{\left|\begin{array}{l}{#1}\\[1ex]{#2}\end{array}\right.}

\begin{document}

\title[A reaction-subdiffusion model of FRAP]{A reaction-subdiffusion model of fluorescence recovery after photobleaching (FRAP)}

\author{S B Yuste $^1$, E Abad $^2$, and K Lindenberg $^3$}

\address{$^1$ Departamento de F\'{\i}sica and Instituto de Computaci\'on Cient\'{\i}fica Avanzada (ICCAEX), Universidad de Extremadura, E-06071 Badajoz, Spain.}
\address{$^2$ Departamento de F\'{\i}sica Aplicada and Instituto de Computaci\'on Cient\'{\i}fica Avanzada (ICCAEX), Centro Universitario de M\'erida, Universidad de Extremadura, E-06800 M\'erida, Spain.}
\address{$^3$ Department of Chemistry and Biochemistry, and BioCircuits Institute, University of California San Diego, 9500 Gilman Drive, La Jolla, CA 92093-0340, USA.}

\begin{abstract}
Anomalous diffusion, in particular subdiffusion, is frequently invoked as a mechanism of motion in dense biological media, and may have a significant impact on the kinetics of binding/unbinding events at the cellular level. In this work we incorporate anomalous diffusion in a previously developed model for FRAP experiments. Our particular implementation of subdiffusive transport is based on a continuous time random walk (CTRW) description of the motion of fluorescent particles, as CTRWs lend themselves particularly well to the inclusion of binding/unbinding events. In order to model switching between bound and unbound states of fluorescent subdiffusive particles, we derive a fractional reaction-subdiffusion equation of rather general applicability. Using suitable initial and boundary conditions, this equation is then incorporated in the model describing two-dimensional kinetics of FRAP experiments. We find that this model can be used to obtain excellent fits to experimental data. Moreover, recovery curves corresponding to different radii of the circular bleach spot can be fitted by a single set of parameters. While not enough evidence has been collected to claim with certainty that CTRW is the underlying transport mechanism in FRAP experiments, the compatibility of our results with experimental data fuels the discussion as to whether normal diffusion or anomalous diffusion is the appropriate model, and as to whether anomalous diffusion effects are important to fully understand the outcomes of FRAP experiments. On a more technical side, we derive explicit analytic solutions of our model in certain limits.

\end{abstract}

\vspace{2pc}
\maketitle

\section{Introduction}
In this work we propose to address the dynamics underlying fluorescence recovery after photobleaching (FRAP), a widely used experimental method to explore binding interactions in cells both in vitro and in vivo. There are of course myriads of papers on the subject, but we will focus on a question that seems not to have been researched. Our focus will be on binding reactions extensively considered both theoretically and experimentally by Sprague et al. in Ref.~\cite{Sprague2004}.

To reveal our particular question, we begin by pointing out a contradiction.  On the one hand, anomalous diffusion (in the form of subdiffusion) is the most common understanding of motion of constituents in crowded media such as, for example, biological cells.  Anomalous diffusion is usually established by considering the mean square displacement of the component of interest, which may, for instance, be   a protein or a portion of a DNA strand or any other component in the crowded cell.  The mean square displacement  of a randomly moving constituent starting from an initial location $\mathbf{r_0}$ is understood to be an average (indicated by brackets) over repeated realizations (measurements) of the motion.  If the mean square displacement $\langle (\mathbf{r} -\mathbf{r_0})^2\rangle$ grows with time $t$ as $t^1$ then the diffusion is ``normal".  On the other hand, if it grows more slowly, as $t^\gamma$ with $\gamma<1$, then the motion is ``anomalous",  specifically ``subdiffusive" because the entity moves more slowly than a normally diffusing one.  (If $\gamma>1$ the motion is ``superdiffusive", a case that has received considerably less attention and that we will not consider in this paper.) Subdiffusion would seem to be a natural description of motion in crowded biological media, and it is so pervasive a description that some authors have begun to question this universal view.  A quote from this latter camp, taken from Ref.~\cite{Dix2008}, goes as follows: ``\emph{We conclude that the notion of universally anomalous diffusion in cells as a consequence of molecular crowding is not correct and that slowing of diffusion in cells is less marked than has been generally assumed.}" Still, in most situations crowding in cells leads to subdiffusion.

The contradiction arises because at the same time that subdiffusion is the model of choice when it comes to motion, theoretical models of FRAP experiments almost always rely on normal diffusion of the binding-unbinding components! This is the problem we wish to address in this work: we propose a complete model in which the reacting species move subdiffusively, and compare the predictions of this model with the experimental results of Sprague et al.~\cite{Sprague2004} and with their model which assumes  normal diffusion. They study the problem of transcription factor mobility. In particular, they measure the FRAP recovery curve of a GFP-tagged glucocorticoid receptor within nuclei of mouse adenocarcinoma cells and compare it extensively with their reaction-diffusion model.

A word about modeling subdiffusion and also including reactions in such models is in order.  There are a number of different models of subdiffusion in the literature.  Two very recent reviews can be found in \cite{Sokolov2012} and \cite{Hofling2013}, and a brief discussion of the difficulties in choosing one particular model can be found in~\cite{Boon2012} . The different ways of modeling subdiffusion lead to clearly distinct macroscopic results for some quantities, but to equivalent results for others.  In the latter case there is therefore no macroscopic basis to distinguish among models. In addition, in many situations one has insufficient knowledge of the microscopic details of transport. All of these issues lead to ongoing debate about which model to use.  It is quite possible that even in a given environment different components move in different ways, or that the description of the motion of a given component is different on different time scales, so that each model may be appropriate under appropriate circumstances.  The inclusion of reactions in the various models brings with it an additional set of uncertainties and issues that lead to even more heated debate.

Our model of choice is based on a continuous time random walk (CTRW). It is our model of choice for two reasons: (1) It is a model that we have worked with for many years in many different contexts, so we are very familiar with it, and (2) It is, to our knowledge, the only model in which chemical reactions have been included analytically, so if we are to describe subdiffusive motion of entities that can undergo binding and unbinding reactions, this is the model that currently allows this description.  We do not claim that this is necessarily \emph{the} correct model, but it is a model of subdiffusion that allows the inclusion of chemical reactions. More forcefully in support of this model, in Ref.~\cite{Barkai2012}, Barkai et al. cite a number of papers that confirm the validity of a CTRW model to explain anomalous diffusion results in a number of biological systems. We stress that the inclusion of reactions in reaction-subdiffusion models is far from trivial, and much more difficult than in normal diffusion, where reaction terms are simply added to the diffusion equation. Simple addition is not appropriate in a subdiffusive model, a point that is central to our discussion presented in detail in the next section.

There are FRAP experiments in biological media that  may need to be described by a subdiffusion-reaction model.  Among them are the binding reaction experiments of Sprague et al.~\cite{Sprague2004}. We choose this system to analyze because the work of Sprague et al. also includes an extensive theoretical analysis based on a reaction-diffusion model that we can now extend to the case of subdiffusion described by a CTRW model.
In Sec.~\ref{sec2} we present a number of necessary definitions and construct a reaction-subdiffusion equation following the approach of Refs.~\cite{Yuste2008a,Yuste2010a}, but now reformulated for the inclusion of a chemical reaction that causes the loss and gain of a species $A$, $A\rightleftarrows 0$. Both death ($A\to 0$) and birth ($0\to A$) contributions are
necessary if the model is to include both binding and unbinding reactions.  In Sec.~\ref{sec3} we use this as a starting point to derive an equation to describe the FRAP recovery. In Sec.~\ref{sec4} we solve the equation for the time Laplace transform of the FRAP recovery curve and find the time-dependent solutions by numerical inversion.  We compare our curves with experimental results obtained by digitalizing the results
in~\cite{Sprague2004} and also with the results of their reaction-diffusion model.  In all of these comparisons it is necessary to make decisions about which parameters to optimize for these comparisons, a choice we discuss in that section. A simplified model initially studied by Sprague et al.~\cite{Sprague2004} and subsequently extended by Lubelski and Klafter~\cite{Lubelski2008} to the CTRW case (our model of choice) is the so-called ``pure-diffusion dominant model."  This is appropriate when most of the fluorescent molecules are free, so that the equation to deal with is a diffusion or subdiffusion equation without a reaction.  We implement the same approximation in Sec.~\ref{sec5}, and show that in this limit it is possible to obtain an analytic time-dependent solution for the FRAP recovery curve, albeit a solution difficult to deal with because it is a complicated function (a Fox H-function, \cite{Mathai1978}).  However, we are able to obtain more transparent expressions for short times and for long times.  Finally, in Sec.~\ref{sec6} we conclude with some final remarks.

\section{ Subdiffusion-reaction equation}
\label{sec2}

The inclusion of reactions in models describing subdiffusive motion is a notoriously difficult problem, far more difficult than in the case of ordinary diffusion \cite{Mendez2010}.  To include reactions in the case of ordinary diffusion involves the simple addition of reaction terms to the diffusion equation.  Such a simple addition in general does not work when the motion is subdiffusive.   In particular, the reaction and subdiffusion contributions affect one another in a complex way.  Furthermore, the model used to describe the subdiffusive motion profoundly affects how reactions enter the problem and, in fact, for most models (fractional Brownian motion, percolation, etc.) this combination has not been considered analytically.  The most extensive work on the problem has been carried out for subdiffusion described as a continuous time random walk (CTRW), and even here the form of the subdiffusion-reaction equation depends on the microscopic description of the way that walkers appear and disappear as they move.  Several extensive references have discussed the problem \cite{Yuste2010a,Mendez2010,Vlad2002,Seki2003,Seki2003a,Yadav2006,Henry2006,
Sokolov2006,Seki2007,Froemberg2008,Henry2008,Fedotov2010a,Yuste2012book,Soula2013,Shkilev2014,Soula2014} and in them one can see that there is no single equation at the mesoscopic description level, as there is for a normal reaction-diffusion problem.

We have focused on a particular description in our work \cite{Yuste2010a} (model B in \cite{Mendez2010,Fedotov2010a}), and it is this description that we use here.  We do not claim that this is ``the correct description" for any particular physical system (although it may hopefully be correct for some), but we are able to offer a complete reaction-subdiffusion model for the FRAP problem that may provide useful insights because it is, to a large extent, analytic.

Before stating the principal features of this model, we recall that in a CTRW there is a waiting time at each location, chosen from a distribution $\psi(t)$, at the end of which a walker takes a step.
Waiting time distributions that give rise to subdiffusion have long time tails that imply that there are often long waiting times between steps.
Typically, the long-time behavior of $\psi(t)$ is of the form
\begin{equation}
\psi(t)\sim \gamma\, t_0^\gamma \: t^{-(1+\gamma)},
\label{waitingtime}
\end{equation}
where $t_0$ is a constant that has the dimension of time. The small $p$ behavior of the Laplace transform $\tilde{\psi}(p)$ is then
\begin{equation}
\tilde{\psi}(p) \sim 1-(\tau_\gamma \, p)^\gamma,
\end{equation}
where $\tau_\gamma = [\Gamma(1-\gamma)^{1/\gamma}\,t_0$ and $\Gamma$ is the Gamma function.
When $\gamma=1$ the mean time between jumps, $\int_0^\infty t\, \psi(t) dt$,  is finite and the walk is diffusive (normal).  Subdiffusion is associated with an exponent $\gamma<1$, which yields an infinite mean time between jumps. The fact that the same waiting time is used for each step means that a clock carried by a walker to measure these events is reset to zero at each step. When reactions are also present, decisions must be made about the timing of birth and death events (only while waiting? only while stepping? at any time?) and about setting the clock at the time of birth of a new particle. Different assumptions lead to different equations.
There is also a distribution $\omega(\mathbf{r})$ that governs the jump lengths and directions.
We restrict ourselves to jump length distributions with finite moments.

The model to be used here has three main features:
\begin{enumerate}
\item  The reaction rate of particles at a given location is proportional to their number at that location.  This is the usual assumption associated with the law of mass action, and is the one used here whether the diffusion is normal or anomalous.
\item Reactions occur at any time, independent of the status of the particles (still or stepping).
\item  Newborn particles as a result of a reactive event are assigned a clock set to zero at the time of birth.  As a result, this model does not distinguish between a particle's appearance at a location by a jump or by a reaction.
\end{enumerate}

This model to describe subdiffusion-reaction problems has been adopted by a number of authors in a variety of contexts. In Refs.~\cite{Yuste2008a,Yuste2010a} we constructed the associated reaction-subdiffusion equations when there are no birth events.  However, the inclusion of these events is essential if we are to describe FRAP.  We thus present here our construction of the corresponding reaction-subdiffusion equations for our model augmented by such birth events.

To construct our reaction-subdiffusion equation based on the CTRW with the features mentioned above, we need to introduce a number of quantities;
\begin{itemize}
\item $c(\mathbf{r},t) = $ concentration of particles,
\item $k(\mathbf{r},t) = $ reaction rate coefficient or reactivity, later taken to be independent of position and time,
\item $j_B(\mathbf{r},t) = $ rate at which the reaction gives birth to new particles,
\item $j(\mathbf{r},t) = $ incoming flux of particles due to the CTRW,
\item $i(\mathbf{r},t) = $ outgoing flux of particles due to the CTRW,
\item $j_T(\mathbf{r},t) = j(\mathbf{r},t) + j_B(\mathbf{r},t) = $ total flux of incoming particles.

\end{itemize}

When normally diffusive particles react, the standard reaction-diffusion equation that describes
the space-time evolution of their concentrations is given by the normal
diffusion equation plus a reaction term, say $F(c)$, that takes into account the rate
of change of $c(\mathbf{r},t)$ due to reactions,
\begin{equation}
\frac{\partial}{\partial t} c(\mathbf{r},t) = D \nabla^2  c(\mathbf{r},t) + F(c).
\end{equation}
However, when particles that diffuse anomalously react, the corresponding
subdiffusion-reaction equation for the concentration is no longer a simple sum~\cite{Vlad2002,Seki2003,Seki2003a,Yadav2006,Henry2006,Sokolov2006,Seki2007,
Froemberg2008,Henry2008,Fedotov2010a}. An extensive and recent discussion of this general topic can be found in \cite{Mendez2010}.

We start with a general description of the gain and loss of particles due to the reaction:
\begin{itemize}
\item The \emph{loss} of
particles at location $\mathbf{r}$ due exclusively to reactions is given by
\begin{equation}
\label{dotc}
\left. \frac{\partial}{\partial t} c(\mathbf{r},t)\right|_{\rm loss-reaction} = -k(\mathbf{r},t)c(\mathbf{r},t).
\end{equation}
\index{reactivity}
\item The \emph{gain} of
particles at location $\mathbf{r}$ due exclusively to reactions is
\begin{equation}
\left. \frac{\partial}{\partial t} c(\mathbf{r},t)\right|_{\rm gain-reaction} = j_{B}(\mathbf{r},t).
\end{equation}
\end{itemize}
We will use these relations in constructing an evolution equation for the concentration of particles as a function of position and time.

To arrive at this evolution equation we build it by carefully combining all the contributions due to jumping and reactions. We start by using Eq.~\eqref{dotc}, that is, by first
considering the situation where the \emph{only} ongoing process is the loss of particles due to the loss reaction. For the moment we set aside the gain due to the reaction as well as the walk.  Integrating Eq.~\eqref{dotc} leads to
 \begin{equation}
\frac{ c(\mathbf{r},t^\prime)}{c(\mathbf{r},t)}\equiv  A(\mathbf{r},t,t^\prime) =
 \exp \left( - \int_{t^\prime}^t k(\mathbf{r},t^{\prime\prime})\,
dt^{\prime\prime}\right),
\end{equation}
which describes the time evolution of the concentration at location $\mathbf{r}$ as time proceeds from $t^\prime$ to $t$ due to the reaction loss. Note that $A(\mathbf{r},t,t^\prime)=[A(\mathbf{r},t^\prime,t)]^{-1}$.

Next we set aside the reaction for a moment and consider
the incoming
and outgoing fluxes of \emph{jumping}  particles at location $\mathbf{r}$ at time $t$. These two fluxes  are  related by the equation
\begin{equation}
 j(\mathbf{r},t) = \int i(\mathbf{r}-\mathbf{r}\,',t)\, \omega(\mathbf{r}\,' )\, d\mathbf{r}\,',
 \label{ij}
\end{equation}
which simply states that the incoming flux of jumping particles at $\mathbf{r}$ at time $t$ arises from the
outgoing fluxes of jumping particles at all other locations  $\mathbf{r}-\mathbf{r}\,'$ at that time.

We are now ready to include all of the contributions to the change in the concentration:
\begin{eqnarray}
\frac{\partial}{\partial t} c(\mathbf{r},t) &= j(\mathbf{r},t) -i(\mathbf{r},t)
-k(\mathbf{r},t)c(\mathbf{r},t)+j_B(\mathbf{r},t)\nonumber\\
&= j_T(\mathbf{r},t) -i(\mathbf{r},t)
-k(\mathbf{r},t)c(\mathbf{r},t).
 \label{balance1}
\end{eqnarray}
This is simply a descriptive statement of the fact that
the changes in the  concentration at $\mathbf{r}$ are due to the incoming
and outgoing fluxes and to the reaction process, both gain and loss, at that location.  It is not yet in the form of an equation to solve for the concentration.

An additional relation connecting the fluxes and concentrations is
\begin{eqnarray}
 i(\mathbf{r},t) &=& \psi(t) A(\mathbf{r},t,0) c(\mathbf{r},0)  + \int_0^t \psi(t-t')A(\mathbf{r},t,t') j_T(\mathbf{r},t')\, dt',
\label{balance2}
\end{eqnarray}
which states that the outgoing flux from $\mathbf{r}$ at time $t$ arises from two sources.  One is
the contribution of the particles that started out at $\mathbf{r}$ at
time $t=0$, did not react or move anywhere up to time $t$,
and then took a step away from $\mathbf{r}$ at time $t$. The other is from those particles
that arrived at $\mathbf{r}$ by a jump or by birth from the reaction
at some earlier time $t'$, waited there up to time $t$ without
degradation, and then stepped away.  Note that as featured above in point (iii), we make no distinction between particles that arrive at a site at a given time due to a jump or due to a reactive gain event. Equations~\eqref{ij}, \eqref{balance1} and \eqref{balance2} together provide a full mathematical description of the problem.  We now proceed to combine them into a single equation. To do so, we need to introduce some additional definitions that will allow us to combine these equations into a convenient form.

We denote the Fourier transform with respect to $\mathbf{r}$ by the symbol $\mathcal{F}$,
and the inverse Fourier transform by $\mathcal{F}^{-1}$.  The Laplace transform with respect to $t$ is denoted by $\mathcal{L}$,
and the inverse Laplace transform by $\mathcal{L}^{-1}$.
The Fourier transform of $w(\mathbf{r})$ is $\hat{w}(\mathbf{q})$ and the Laplace transform of $\psi(t)$ is $\tilde{\psi}(p)$.
Next we introduce the Gr\"unwald-Letnikov fractional time derivative $~_{0}\mathcal{D}_t^{1-\gamma}$ whose Laplace transform is~\cite{Podlubny1999}
\begin{equation}
\mathcal{L}~_{0}\mathcal{D}_t^{1-\gamma} f(t) = p^{1-\gamma} \tilde{f}(p).
\label{RL}
\end{equation}
When operating on sufficiently smooth functions $f$
(functions $f(t)$ for which $\lim_{t\to 0} \int_0^t d\tau(t-\tau)^{1-\gamma} f(\tau) = 0)$,
this operator is equivalent to the Riemann-Liouville fractional derivative~\cite{Podlubny1999}
\begin{equation}
~_{0}D_t^{1-\gamma} f(t)
=\frac{1}{\Gamma(\gamma)}\frac{\partial}{\partial t}
\int_0^t dt'\, \frac{f(t')}{(t-t')^{1-\gamma}}.
\end{equation}
Finally, we introduce the Riesz fractional spatial derivative $\nabla^\mu$ whose Fourier transform is
\begin{equation}
\mathcal{F} \nabla^\mu g(\mathbf{r})  = -q^\mu \hat{g}(\mathbf{q})
\label{Fourier}
\end{equation}
for sufficiently smooth functions $g$~\cite{Kilbas2006}. In this work we set $\mu=2$. The spatial derivative $\nabla^2$ is the Laplacian, which is the derivative that enters the problem when the jump distribution has a finite second moment.

We now return to our effort to combine our three equations into a single one for $c(\mathbf{r},t)$.
Defining $c^*(\mathbf{r},t)\equiv c(\mathbf{r},t)A(\mathbf{r},0,t)$, we can write
Eq.~\eqref{balance1} as
\begin{eqnarray}
 A(\mathbf{r},t,0)\frac{\partial}{\partial t} c^*(\mathbf{r},t) &= j_T(\mathbf{r},t) -i(\mathbf{r},t)\nonumber\\
 &= j(\mathbf{r},t) -i(\mathbf{r},t)+j_B(\mathbf{r},t).
 \label{balance1a}
\end{eqnarray}
From  Eq.~(\ref{ij}) it then follows that
\begin{equation}
A(\mathbf{r},t,0)  \frac{\partial}{\partial t} c^*(\mathbf{r},t)=
\mathcal{F}^{-1}
\left\{\left[\hat{\omega}(\mathbf{q})-1\right]\hat{i}(\mathbf{q},t)\right\} + j_B(\mathbf{r},t).
\end{equation}
In the diffusive limit ($|q|\to 0$) and for symmetric step-size distributions ($\langle \omega\rangle =0$), one has $\hat{\omega}(\mathbf{q})-1 = -\sigma^2 q^2$, and then
\begin{equation}
 A(\mathbf{r},t,0) \frac{\partial}{\partial t} c^*(\mathbf{r},t) = \sigma^2 \nabla^2 i(\mathbf{r},t)+j_B(\mathbf{r},t),
\label{almost1}
\end{equation}
where $2\sigma^2$ is the variance of the step length distribution.
Upon Laplace transforming  Eq.~(\ref{balance2}) with respect to time
one finds that
\begin{equation}
 \mathcal{L}\left[A(\mathbf{r},0,t) i(\mathbf{r},t)\right]=\tilde{\psi}(p) c^*(\mathbf{r},0) +
\tilde{\psi}(p)  \mathcal{L}\left[A(\mathbf{r},0,t) j_T(\mathbf{r},t)\right].
 \label{icj*}
\end{equation}
Taking into account that, from Eq.~\eref{balance1},
\begin{equation}
 \mathcal{L}\left[A(\mathbf{r},0,t) j_T(\mathbf{r},t)\right]=   \mathcal{L} \left[\frac{d c^*}{dt} \right]+  \mathcal{L}\left[A(\mathbf{r},0,t) i(\mathbf{r},t)\right],
\end{equation}
Eq.~\eref{icj*} can be rewritten as
\begin{equation}
 i(\mathbf{r},t) = A(\mathbf{r},t,0)\mathcal{L}^{-1} \left\{
\frac{p\tilde{\psi}(p)}{1-\tilde{\psi}(p)} \tilde{c}^*(\mathbf{r},p)\right\}.
\end{equation}
Using the Laplace transform expression  of the fractional derivative
operator, this expression can be reformulated as
\begin{equation}
 i(\mathbf{r},t) = A(\mathbf{r},t,0)\tau^{-\gamma}\, _{0}\mathcal{D}_t^{1-\gamma} c^*(\mathbf{r},t).
 \label{almost2}
\end{equation}
Finally, inserting Eq.~(\ref{almost2}) into Eq.~(\ref{almost1}) and expanding
the abbreviated notation, we arrive at the general reaction-subdiffusion \index{reaction-subdiffusion equation}
equation that is the starting point of our analysis:
\begin{eqnarray}
\frac{\partial}{\partial t} c(\mathbf{r},t) &=& D_\gamma \nabla^2   \left\{ e^{-\int_0^t k(\mathbf{r},t^\prime)dt^\prime }~_{0}\mathcal{D}_t^{1-\gamma} \left[
e^{\int_0^t k(\mathbf{r},t^\prime) dt^\prime}c(\mathbf{r},t)\right]\right\} \nonumber\\
&&-k(\mathbf{r},t)c(\mathbf{r},t)+j_B(\mathbf{r},t),
\label{eqdifugral}
\end{eqnarray}
where $D_\gamma=\sigma^2/\tau_\gamma^\gamma$. In this equation it can clearly be seen that the reaction and subdiffusion contributions are not simply added.  The last two terms would be those included in a normal diffusion-reaction equation as the loss portion (penultimate term) and the birth or gain portion (last term).  But the first term on the right is not simply a subdiffusion contribution. It contains the contribution of subdiffusion enmeshed in a complex way with the loss reaction, a way that could not easily have been predicted from pure phenomenology at this level.

In this work we do not consider a
time-dependent reactivity.  Our starting equation therefore is
\begin{eqnarray}
\frac{\partial}{\partial t} c(\mathbf{r},t) &= D_\gamma \nabla^2 \left\{ e^{- k(\mathbf{r})t}~ _{0}\mathcal{D}_t^{1-\gamma} \left[
e^{k(\mathbf{r})t} c(\mathbf{r},t)\right]\right\} -k(\mathbf{r})c(\mathbf{r},t)+j_B(\mathbf{r},t)\nonumber\\
&= D_\gamma  \nabla_R^2  c(\mathbf{r},t) -k(\mathbf{r})c(\mathbf{r},t)+j_B(\mathbf{r},t),
\label{ourstart}
\end{eqnarray}
where we have introduced the subdiffusion-reaction operator
\be
\nabla_R^2\, c(\mathbf{r},t)=\nabla^2 \left\{ e^{- k(\mathbf{r})t}~ _{0}\mathcal{D}_t^{1-\gamma} \left[ e^{k(\mathbf{r})t} c(\mathbf{r},t)\right]\right\}.
\label{nabla2R}
\ee

Finally, in order to solve Eq.~\eqref{ourstart} it is convenient to work with a new function $v(\mathbf{r},t)$ whose Laplace transform is
\be
\label{vxudef}
\tilde v(\mathbf{r},p)=[p+k(\mathbf{r})]^{1-\gamma}\, \tilde c(\mathbf{r},p).
\ee
Taking into account that
\begin{eqnarray}
\mathcal{L} \left[ e^{-k(\mathbf{r}) t}   ~_{0}{\cal D}_t^{1-\gamma}\left( e^{k(\mathbf{r}) t}
c(\mathbf{r},t)\right) \right]= [p+k(\mathbf{r})]^{1-\gamma} \tilde c(\mathbf{r},p),
\end{eqnarray}
Eq.~\eqref{ourstart} becomes
\begin{eqnarray}
\label{cxunu}
p\,\tilde c(\mathbf{r},p)-c(\mathbf{r},0)&=& D_\gamma  \nabla^2
\left\{ [p+k(\mathbf{r})]^{1-\gamma} \tilde c(\mathbf{r},p) \right\}
-k(\mathbf{r}) \tilde c(\mathbf{r},p)+j_B(\mathbf{r},p)\nonumber\\&&
\end{eqnarray}
or, equivalently,
\be
\label{vxunu}
[p+k(\mathbf{r})]^{\gamma} \tilde v(\mathbf{r},p)-c(\mathbf{r},0)= D_\gamma  \nabla^2  \tilde v(\mathbf{r},p)+j_B(\mathbf{r},p).
\ee
This, then, is a variant of our starting equation that describes subdiffusion as well as a reaction that depletes and gives birth to particles.

\section{Subdiffusive FRAP model}
\label{sec3}

The FRAP system considered here is that studied by Sprague et al.~\cite{Sprague2004} (and extended by Mueller et al.~\cite{Mueller2008} to a variety of geometries not considered here), but we generalize their normal diffusion model to the subdiffusive case described by the CTRW.

Freely subdiffusing proteins
undergo transient binding events with immobile nuclear structures.  As a result, there is a concentration $c(\mathbf{r},t)$ of bound proteins, a concentration $s(\mathbf{r},t)$ of vacant binding sites, and a concentration $f(\mathbf{r},t)$ of free proteins.
The reaction then proceeds according to the scheme
\[
\begin{array}{ccl}
& k_{on}&\\
F+S
&\stackrel{\textstyle \longrightarrow}{\longleftarrow}& C\\
& k_{off}&\\ \\
\end{array}
\]
Here $F$ represents free proteins, $S$ denotes vacant binding sites, and $C$ represents bound $[FS]$ complexes.  The rate coefficients $k_{on}$ and $k_{off}$ are  for binding and unbinding, respectively.  This scheme in principle requires us to write three reaction-subdiffusion equations, one for each of the three concentrations.

However, the complexity of the problem is considerably reduced by implementing simplifying assumptions. The first is that the biological system has reached equilibrium before photobleaching. Now, FRAP recovery occurs on time scales of seconds to minutes, while GFP-fusion expression takes much longer, a time scale on the order of hours. Furthermore, the GFP fusion proteins have typically reached a constant level by the time the FRAP experiments begin.  We therefore assume that  before the bleach the system is at equilibrium, with the concentrations of free and bound proteins and of vacant binding sites at their uniform steady-state values $F_{eq}$, $C_{eq}$, and $S_{eq}$. Bleaching changes the number of visible free and complexed molecules, but it does not change the number of free binding sites.  We therefore need not include an equation for $s(\mathbf{r},t)$: this concentration is equal to $S_{eq}$ throughout the experiment.  This reduces the number of equations from three to two.  Reaction rate contributions of the form $k_{on} f(\mathbf{r},t) s(\mathbf{r},t)$ can therefore be replaced by $k_{on} S_{eq} f(\mathbf{r},t)$. The product $k_{on} S_{eq}$ is then a pseudo-first-order rate constant that we denote as $k_{on}^*$.
The concentrations of proteins are normalized so that $F_{eq}+ C_{eq}=1$.

This would then leave us with two subdiffusion-reaction equations, one for $c(\mathbf{r},t)$ and another for $f(\mathbf{r},t)$.  The reaction terms are all of first (or pseudo-first) order.  A further simplification can be made by noting that the binding sites are part of a large complex.  This complex is relatively immobile on the time scale of the FRAP experiment.  We can then eliminate the
subdiffusion part of the equation for $c(\mathbf{r},t)$ and assume that this component is physically stationary.  That finally leaves us with one subdiffusion-reaction equation for $f(\mathbf{r},t)$ and a pure reaction equation for $c(\mathbf{r},t)$.

Our starting equations are then similar to those of Ref.~\cite{Sprague2004}, but complicated by the fact that normal diffusion is now replaced by subdiffusion:
\begin{eqnarray}
\label{evoleqc1}
\frac{\partial}{\partial t} f(\mathbf{r},t) &= D_\gamma \nabla^2 \left\{ e^{- k^*_{on} t}~ _{0}D_t^{1-\gamma} \left[ e^{ k^*_{on} t} f(\mathbf{r},t)\right]\right\} -k^*_{on}\,f(\mathbf{r},t) + k_{off}\, c(\mathbf{r},t)  ,
\nonumber \\
\label{evoleqc2}
\frac{\partial}{\partial t} c(\mathbf{r},t)&=k^*_{on}\, f(\mathbf{r},t) - k_{off} \,c(\mathbf{r},t).
\end{eqnarray}
As said earlier, the system is at equilibrium before the bleach, so that initially $df/dt=dc/dt=0$.  As a result,
\begin{equation}
\frac{F_{eq}}{C_{eq}} = \frac{k_{off}}{k^*_{on}} .
\label{4S}
\end{equation}
The normalization $F_{eq}+ C_{eq}=1$ leads to the values
\begin{equation}
F_{eq}=\frac{k_{off}}{k^*_{on}+k_{off}},
\qquad
C_{eq}=\frac{k^*_{on}}{k^*_{on}+k_{off}}.
\label{5bS}
\end{equation}
Next, at the site of the bleach the concentration of fluorescent molecules is reduced by photobleaching, and the return to equilibrium is dictated by Eqs.~\eqref{evoleqc2}. The measured FRAP recovery data is then the sum of free and bound fluorescence averaged over the bleach spot: $frap(t)= \langle f(t)\rangle + \langle c(t)\rangle$, where the brackets $\langle \cdots \rangle$ denote the spatial average.
The steps involved in averaging the experimental results are discussed in ~\cite{Sprague2004}.
The assumption that the initial concentrations are indeed the equilibrium concentrations relies on the fact that the photobleach is fast.  The assumption that the concentrations return to the initial normalized equilibrium values relies on the bleach spot being small relative to the total cell volume because otherwise some non-negligible fraction of fluorescence would be lost after the bleach. Finally, we assume, along with most other theoretical FRAP work, that diffusion takes place only in two dimensions, in the plane of focus.  This is appropriate when the bleaching area forms an essentially cylindrical shape through the cell, as is usually the case ~\cite{Sprague2004}. The axial terms in the equations of motion then do not need to be included in the Laplacian $\nabla^2$, and only the radial  components remain.
We finally note that, for the sake of simplicity, it is important to avoid aging effects characteristic of CTRW models  \cite{Sokolov2012,Lubelski2008}.



We introduce the transformation
\be
u(\mathbf{r},t) =F_{eq}-f(\mathbf{r},t),\qquad v(\mathbf{r},t)=C_{eq}-c(\mathbf{r},t).
\ee
It is straightforward to establish from the above initial conditions that
\begin{equation}
\frac{u(\mathbf{r},0)}{v(\mathbf{r},0)}=\frac{k_{off}}{k_{on}^*}.
\label{uvkk}
\end{equation}
The evolution equations for $u$ and $v$ are directly found to be
\begin{eqnarray}
\frac{\partial}{\partial t} u(\mathbf{r},t) &= D_\gamma \nabla^2_R   u(\mathbf{r},t)  -k^*_{on}\,u(\mathbf{r},t) + k_{off}\, v(\mathbf{r},t),
\\
\frac{\partial}{\partial t} v(\mathbf{r},t)&=k^*_{on}\,u(\mathbf{r},t) - k_{off} \,v(\mathbf{r},t) .
 \end{eqnarray}
These equations are similar to Eq.~(12) in Ref.~\cite{Sprague2004} except for the important replacement of $\nabla^2$ in the case of normal diffusion by our considerably more complicated operator $\nabla_R^2$ defined in Eq.~\eqref{nabla2R}.

The evolution equations are most readily solved by first Laplace transforming them with respect to time:
\begin{eqnarray}
 p \tilde u(\mathbf{r},p) &= D_\gamma \left(p +k^*_{on} \right)^{1-\gamma}  \nabla^2 \tilde u(\mathbf{r},p)
 - k^*_{on} \tilde u(\mathbf{r},p) + k_{off} \tilde v(\mathbf{r},p) +u(\mathbf{r},0) ,
 \label{13aSbis}\\
p \tilde v(\mathbf{r},p) &=   k^*_{on} \tilde u(\mathbf{r},p) - k_{off} \tilde v(\mathbf{r},p) +v(\mathbf{r},0).
\end{eqnarray}
From the second equation it immediately follows that
\begin{eqnarray}
\label{vtil}
 \tilde v(\mathbf{r},p) =   \frac{k^*_{on} \tilde u(\mathbf{r},p)  +v(\mathbf{r},0)}{p+k_{off}}.
\end{eqnarray}
When this is substituted back into Eq.~\eqref{13aSbis}, we obtain an equation for the single remaining as yet unknown function $\tilde{u}(r,p)$.
Subsequently we recognize that the radially symmetric initial condition means that the solutions as time evolves are also radially symmetric, that is, all $\mathbf{r}$ dependences are in fact dependences on $r\equiv |\mathbf{r}|$.

Since the FRAP recovery is the sum of the free ($f=F_{eq}-u$)  and bound
fluorescence ($c=C_{eq}-v$) , we must compute the Laplace transform for this sum, $f+c=1-u-v$. This yields the Laplace transform of the fluorescence intensity as a function of radial position within the bleach spot as
\begin{eqnarray}
\widetilde{fluor}_\gamma(r,p)&=\frac{1}{p}-\tilde u(r,p)-\tilde v(r,p).
\label{vtil2}
\end{eqnarray}
Substituting Eq.~\eqref{vtil} into Eq.~\eqref{vtil2} then yields
\begin{eqnarray}
\widetilde{fluor}_\gamma(r,p)&=\frac{1}{p}-\tilde u(r,p) \left(1+\frac{k^*_{on}}{p+k_{off}}\right)-
\frac{C_{eq}}{p+k_{off}}.\\ \nonumber
\end{eqnarray}

To obtain the measured FRAP recovery, we must compute the average
fluorescent intensity within the measurement region of radius $w$:
\begin{eqnarray}
\widetilde{frap}_\gamma(p)&=
 \langle \widetilde{fluor}(r,p)\rangle\\
 &=\frac{1}{p}-\langle \tilde u(r,p) \rangle-\langle \tilde v(r,p) \rangle,
  \end{eqnarray}
that is,
\begin{eqnarray}
\widetilde{frap}_\gamma(p) =
\frac{1}{p}-\langle \tilde u(r,p) \rangle  \left(1+\frac{k^*_{on}}{p+k_{off}}\right)-
\frac{C_{eq}}{p+k_{off}}.
\label{fraps}
  \end{eqnarray}
To determine the Laplace transform of the FRAP function recovery it is thus only necessary to calculate $\langle \tilde{u}(r,p)\rangle$:
\begin{eqnarray}
 \langle \tilde {u}(r,p) \rangle= \frac{1}{\pi w^2} \int_0^{2\pi} d\theta \int_0^{w} dr\, r\, \tilde u(r;p)=
  \frac{2}{ w^2}   \int_0^{w} dr\, r\, \tilde u(r;p).
  \label{uavg2}
\end{eqnarray}

As noted earlier, substitution of Eq.~\eqref{vtil} into \eqref{13aSbis} yields a closed equation for $\tilde{u}(r,p)$:
 \begin{eqnarray}
 0&= D_\gamma \left(p +k^*_{on} \right)^{1-\gamma}  \nabla^2 \tilde{u}(r,p) -p\left(1+ \frac{k^*_{on}}{p+k_{off}}\right) \tilde{u}(r,p)\nonumber\\
&~~~ +   \left(1+ \frac{k^*_{on}}{p+k_{off}}\right) u(r,0)
\end{eqnarray}
or, equivalently,
 \begin{eqnarray}
 \label{usg}
 \nabla^2 \tilde{u}(r,p) - q^2_\gamma \tilde{u}(r,p) =  V_\gamma(r,p),
\end{eqnarray}
where
\begin{eqnarray}
 q^2_\gamma&
 =\frac{p}{D_\gamma (p+k^*_{on})^{1-\gamma}} \left( 1+\frac{k^*_{on}}{p+k_{off}}\right)\\
 V_\gamma&=  -\frac{u(r,0)}{D_\gamma (p+k^*_{on})^{1-\gamma}} \left(1+\frac{k^*_{on}}{p+k_{off}}\right).
\end{eqnarray}
Our task is  then to find the solution of Eq.~\eqref{usg} that also satisfies the boundary conditions.  We proceed to do so in the next section.

\section{Uniform circular disk model}
\label{sec4}

The equations and simplifications introduced and discussed above are appropriate for an initial bleach spot that can be considered to be a two-dimensional region with radial (i.e., cylindrical) symmetry.  In particular, they are appropriate for a uniform circular disk model of the bleach region in which the initial condition is
\begin{equation}
\label{x}
u(r,0)=
\left\{
\begin{array}{ll}
F_{eq}, & r\leq w \\
0, & r > w
\end{array}
\right.
\end{equation}
As noted earlier, this is the geometry first discussed by Sprague et al.\cite{Sprague2004}.

The difference between our equation for the Laplace transform $\tilde{u}(r,p)$ of $u(r,t)$ as given in Eq.~\eqref{usg} and that of Sprague et al.'s Eq.~(15) lies in the functions $q_\gamma^2$ and $V_\gamma$. Normal diffusion corresponds to the choice $\gamma=1$, which simplifies these functions. However, this simplification does not enter in a practically significant way until we carry out the inverse Laplace transform.  In other words, the solution for the Laplace transform $\tilde{u}(r,p)$ of Sprague et al. is transferable to our problem with the substitutions $q \to q_\gamma$ and $V \to V_\gamma$, where $q$ and $V$ in their notation are $q_1$ and $V_1$ in ours. We therefore refer the reader to the details in~\cite{Sprague2004} (Appendix). Here we just mention the main steps.

This system is of the form seen as far back as the previous mid-century to describe heat conduction, with well-established solutions~\cite{Carslaw1959}.  One finds
\begin{equation}
\label{y}
\tilde{u}(r,p)=
\left\{
\begin{array}{ll}
(V_\gamma/q_\gamma^2) - \alpha_1 I_0(q_\gamma r) \quad &r\leq w _{eq}, \\
 \alpha_2 K_0(q_\gamma r) \quad \qquad& r>w,
\end{array}
\right.
\end{equation}
where $I_0$ and $K_0$ are modified Bessel functions of the first and second kind, respectively. The constants $\alpha_1$ and $\alpha_2$ are determined by requiring that $\tilde{u}$  and its first derivative with respect to $r$ be continuous across the boundary $r=w$. This condition leads to $\alpha_1=(V_\gamma/q_\gamma^2)q_\gamma w K_1(q_\gamma w)$, the important constant for our purposes.  Following the sequence of steps presented in the last section then directly leads to
\begin{equation}
    \widetilde{frap}_\gamma(p)=\frac{1}{p}-\frac{F_{eq}}{p}
    \left[  1-2 K_1(q_\gamma w) I_1(q_\gamma w) \right]\left(1+\frac{k^*_{on} }{p+k_{off} }
     \right)- \frac{C_{eq} }{p+k_{off} }.
\label{22S}\end{equation}
This agrees with Eq.~(22) of Sprague et al. with the substitutions $q_\gamma \to q_1\equiv q$ and $V_\gamma\to V_1\equiv V$.

Equation~\eqref{22S} can be simplified using the normalization condition $F_{eq}+C_{eq}=1$ together with Eq.~\eqref{5bS}. One easily sees that
\be
\frac{1}{p}-\frac{F_{eq}}{p}\left(1+\frac{k^*_{on} }{s+k_{off} }\right)-\frac{C_{eq} }{p+k_{off} }\equiv 0,
\ee
so that
\be
\widetilde{frap}_\gamma(p)=\frac{2F_{eq}}{p}\left[ K_1(q_\gamma w) I_1(q_\gamma w) \right]\left(1+\frac{k^*_{on} }{p+k_{off} }
     \right).
     \label{ns1}
\ee
This is the result we will continue to use in the remainder of this section.  However, it is worth noting that Eq.~(22) of Sprague et al. for normal diffusion can also be simplified to
\be
\widetilde{frap}_1(p)=\frac{2F_{eq}}{p}\left[ K_1(q w) I_1(q w) \right]\left(1+\frac{k^*_{on} }{p+k_{off} }
     \right).
     \label{ns2}
\ee
Neither Eq.~\eqref{ns1} nor even Eq.~\eqref{ns2} can be Laplace inverted analytically.  To find the time-dependent FRAP curves requires numerical inversion.

To determine whether subdiffusion is as good a model to describe the FRAP process than is ordinary diffusion, or perhaps even better, it is helpful to compare both models to experimental measurements. Experimental results for the uniform circular disk geometry are presented in Fig. 5 of Sprague et al.~\cite{Sprague2004}, obtained for FRAP recovery to nuclear mobility of a green fluorescent protein (GFP)-tagged glucocorticoid receptor (GFP-GR) in nuclei of both normal and ATP-depleted cells.  In some of these figures the data is shown relatively cleanly and can therefore be digitalized fairly easily.

It is particularly helpful that Sprague et al. adjusted a number of parameters to optimize the fit of results obtained from a normal diffusion model.  We can use some of the same parameters in testing subdiffusion instead of attempting to readjust all the parameters. We could of course attempt to optimize all the parameters, but that would be an extensive task and not necessary to make our point.
We use the same values for $k_{on}^*$ as obtained by Sprague et al~\cite{Sprague2004}.  We can of course not directly translate their diffusion coefficient to our subdiffusion problem, so we choose our $\gamma$-dependent $D_\gamma$ as follows.  A characteristic time $\tau_1$ for a normally diffusing walker to cover a disk area of radius $w$ is often defined via the relation $w^2=4 D_1 \tau_1$.  Similarly, in anomalous diffusion models  a characteristic time $\tau_\gamma$ to cover the disk area is frequently defined via the relation $w^2=4 D_\gamma \tau_\gamma^\gamma/\Gamma(1+\gamma)$ (which reduces to the one above when $\gamma=1$).  We choose the times $\tau_1$ and $\tau_\gamma$ to be equal. This seems to us a reasonable way to scale the times with respect to one another in the two problems, given the fact that the radius $w$ of the region and the characteristic time to cover this region, $\tau_1\equiv\tau_\gamma$, are easily measurable quantities.  This then implies a relation between $D_1$ and $D_\gamma$ for arbitrary $\gamma < 1$.  In the end we test several values of $\gamma$.  Once having chosen $\gamma$, the only parameter that we fit so as to optimize the agreement of our model to the experimental results is $k_{off}$.  Our optimal values change with changing $\gamma$ but are in the same range as those obtained by Sprague et al.~\cite{Sprague2004} in the fit of the normal diffusion curves.  We note that we could just as well have chosen the same value of $k_{off}$ as found by Sprague et al. and proceeded to optimize the choice of $k_{on}^*$, or optimize with respect to both, but our final conclusions would not change.

Sprague et al. use a ``full model" (in their terminology, this means that they solve the full reaction-diffusion equations without further approximations than those introduced above) for comparisons with experimental results.  When the radius of the circular disk is $w=1.1 \mu$m, the experimental results shown in Fig.~5E of Sprague et al. are best reproduced by the parameter values $k_{on}^*= 500 s^{-1}$ and $k_{off}=86.4 s^{-1}$.  In Fig.~5F the results are shown for a radius $w= 0.5 \mu$m, and the best fit is obtained with $k_{on}^*= 400 s^{-1}$ and $k_{off}=78.6 s^{-1}$. The diffusion constant is estimated for both radii to be $D_1=9.2\mu$m$^2$/$s$. In their notation, $D_1\equiv D_f$.

\begin{figure}
  \centering
    \includegraphics[width=0.6\textwidth,angle=0]{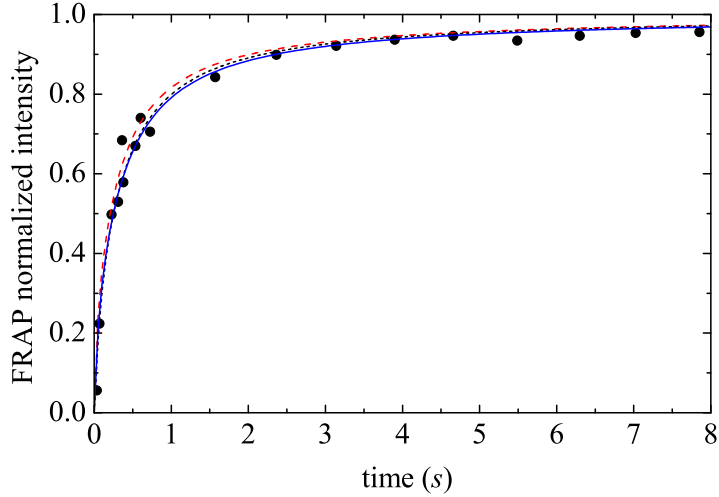}
  \caption{Fig.~5E of Sprague et al. 2004. Circular disk radius: $w=1.1 \mu$m. Dots: experimental results.
Dashed red curve: Normal diffusion ($\gamma=1$). Solid blue curve: CTRW (anomalous diffusion) with $\gamma=0.8$. Dashed black curve: CTRW (anomalous diffusion) with $\gamma=0.5$.  See text for parameter values. }
  \label{Sprague5E}
\end{figure}

\begin{figure}
  \centering
    \includegraphics[width=0.6\textwidth,angle=0]{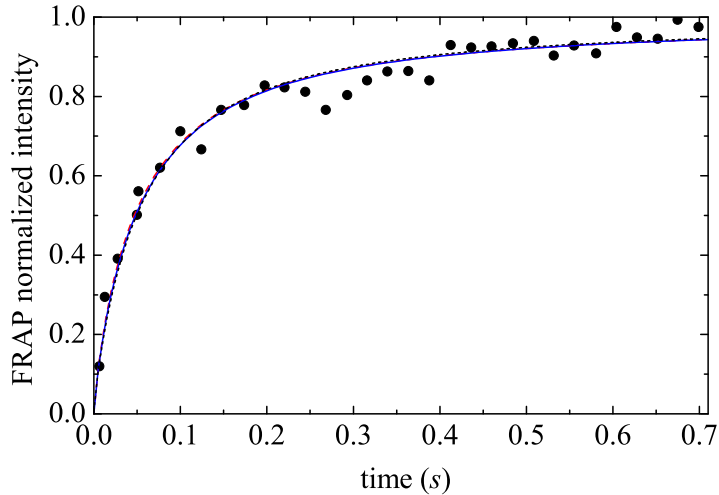}
  \caption{Fig.~5F of Sprague et al. 2004. Circular disk radius: $w=0.5 \mu$m. Dots: experimental results.
Dashed red curve: Normal diffusion ($\gamma=1$). Solid blue curve: CTRW (anomalous diffusion) with $\gamma=0.8$. Dashed black curve: CTRW (anomalous diffusion) with $\gamma=0.5$.  See text for parameter values. }  \label{Sprague5F}
\end{figure}

In Figs.~\ref{Sprague5E} and \ref{Sprague5F}  we show the experimental results and the results of the full reaction-diffusion model of Sprague et al.~\cite{Sprague2004}, along with results obtained from our CTRW formalism.  The Sprague et al. results are shown in both figures by dots (experimental results) and by a black curve (full reaction-diffusion results).

In Fig.~\ref{Sprague5E} we show CTRW curves for the radius $w=1.1 \mu$m.  The parameters are as follows:
\begin{itemize}
\item   $k_{on}^*=500 \, s^{-1}$,  $k_{off}= 86.4 , s^{-1}$, $D_f\equiv D_1=9.2\, \mu m^2/s$, $\gamma=1$.
\item  $k_{on}^*=500 \, s^{-1}$,  $k_{off}= 42.4 , s^{-1}$, $D_\gamma=4.3\, \mu $m$^2/s^\gamma$, $\gamma=0.8$.
\item  $k_{on}^*=500 \, s^{-1}$,  $k_{off}= 20.3 , s^{-1}$, $D_\gamma=1.5\, \mu $m$^2/s^\gamma$, $\gamma=0.5$.
\end{itemize}
in Fig.~\ref{Sprague5F}  we show CTRW curves for the radius $w=1.1 \mu$m.  The parameters are as follows:
\begin{itemize}
\item   $k_{on}^*=400 \, s^{-1}$,  $k_{off}= 78.6 , s^{-1}$, $D_f\equiv D_1=9.2\, \mu m^2/s$, $\gamma=1$.
\item  $k_{on}^*=400 \, s^{-1}$,  $k_{off}= 68 , s^{-1}$, $D_\gamma=3.2\, \mu $m$^2/s^\gamma$, $\gamma=0.8$.
\item  $k_{on}^*=400 \, s^{-1}$,  $k_{off}= 54.7 , s^{-1}$, $D_\gamma=0.7\, \mu $m$^2/s^\gamma$, $\gamma=0.5$.
\end{itemize}
The fact that the lines corresponding to each set of parameters fall on top of each other and capture the experimental points shows that the CTRW provides as compatible a description of the experiments as does normal diffusion.  There is thus no way at this point to choose one over the other.  Note that the fits are very good even for the rather strongly anomalous case $\gamma=0.5$.

\begin{figure}
  \centering
    \includegraphics[width=0.6\textwidth,angle=0]{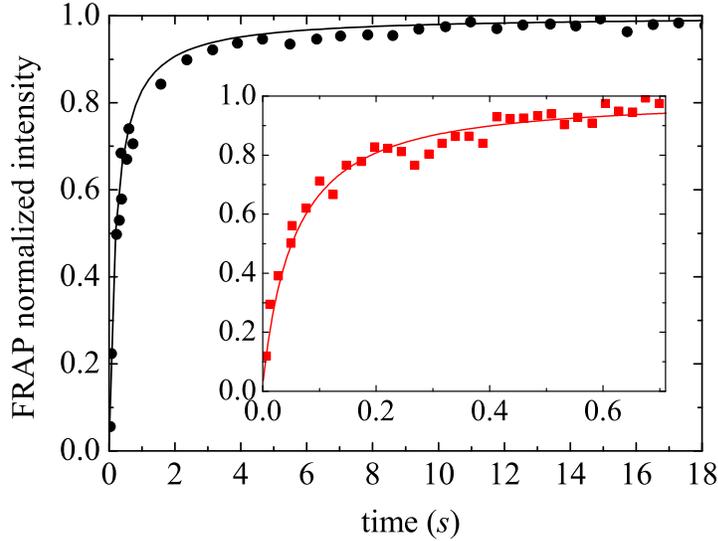}
  \caption{Experimental results for $w=1.1\mu$m (black) and $w=0.5\mu$m (inset, red).  Curves are for $\gamma=0.8$ in both cases.  The common values of the parameters for both figures are $D_\gamma = 4.0\, \mu $m$^2/s^\gamma$, $k_{on}^*=500s^{-1}$, and $k_{off}=60s^{-1}$.}
 \label{ParaComun}
\end{figure}

It was already noted by Sprague et al., and we note again here, that one would not expect $k_{on}^*$ and $k_{off}$ to change with changing radius $w$ of the bleach area. However, we see that the optimal fit in each case does involve a change in $k_{off}$ when we keep $k_{on}$ fixed. We see that the ratio $k_{off}/k_{on}^*$ decreases when $\gamma$ decreases.  It is remarkable that this behavior agrees with the one predicted by Shkilev~\cite{Shkilev2014} by means of a random trap model that by this measure mimics the CTRW model considered here. On the other hand, we find that it is in fact possible to choose common parameters as $w$ changes, and this is shown in Fig.~\ref{ParaComun}.  Here we show experimental results and curves for the anomalous diffusion case with $\gamma=0.8$, with common parameters for the radii $w=1.1 \mu$m and $w=0.5 \mu$m. The common values used in the figure, which shows the results for $w=1.1 \mu$m in red and $w=0.5 \mu$m in blue, are $D_\gamma = 4.0\, \mu $m$^2/s^\gamma$, $k_{on}^*=500s^{-1}$, and $k_{off}=60s^{-1}$.

In spite of the excellent fits for two different values of $w$ exhibited in Fig.~\ref{ParaComun} with a single set of parameter values, we wish to stress that the theoretical results are quite sensitive to parameter choices.  It should also be noted that there exist other different pairs of $k_{on}^*, k_{off}$ that lead to similarly good fits. This behavior is in fact also found for normal diffusion~\cite{Mueller2008}.

\section{Reduction of the full model to the pure-subdiffusion dominant model}
\label{sec5}

The full model, even in the case of normal diffusion but also in the case of anomalous diffusion, reduces to simpler forms in some limiting cases. Three particular cases are considered by Sprague et al.~\cite{Sprague2004}. The first is called ``pure-diffusion dominant" and arises when most of the fluorescent molecules are free.  FRAP then measures mainly free diffusion of these fluorescently tagged molecules. The second, called ``effective diffusion," arises when the reaction is much faster than diffusion.
The third is the ``reaction dominant" case, when diffusion is very fast compared to both binding and to the timescale of the FRAP measurement. We briefly discuss the first of these cases for the CTRW problem.
In the case of normal diffusion, the second leads to an equation similar to that of the first, but with a modified diffusion coefficient, $D_{1,eff}=D_1/\left[1+(k_{on}^*/k_{off})\right]$.  We conjecture a similar result for subdiffusion, with the appropriately modified  subdiffusion coefficient.
The third leads to a FRAP recovery given by $frap(t) = 1-C_{eq}e^{-k_{off}t}$.  Again, we conjecture that the same is seen in the case of subdiffusion provided the subdiffusion is again very fast compared to binding and to the timescale of the FRAP measurement.

In the pure-(sub)diffusion dominant limit we need to solve Eqs.~\eqref{evoleqc2} when $k_{on}^*$ and $k_{off}$ are set equal to zero. The second equation then trivially gives $c(\mathbf{r},t)=C_{eq}$.
In the case of normal diffusion ($\gamma=1$) this leads to the ordinary diffusion equation
\begin{equation}
\frac{\partial}{\partial t}f(\mathbf{r},t) = D_1\nabla^2 f(\mathbf{r},t),
\end{equation}
whose solution in closed form for our geometry is well known.  The FRAP recovery curve obtained as a result by Soumpasis is ~\cite{Soumpasis1983}
\begin{equation}
frap_{1,PD}(t) = e^{-\tau_1/2t}\left[I_0\left(\frac{\tau_1}{2t}\right) + I_1\left(\frac{\tau_1}{2t}\right)\right],
\label{gammais1}
\end{equation}
where the $I_i$ are modified Bessel functions, and
\begin{equation}
\tau_1=w^2/D_1.
\label{tau1}
\end{equation}
In their notation $\tau_1\equiv\tau_D$ and, as always, $D_1\equiv D_f$. The additional subscript $PD$ stresses that this is the pure-diffusion dominant solution.
Sprague et al.~\cite{Sprague2004} further discuss how this solution provides helpful information for the more general situation where the full model is appropriate.

In the subdiffusive case, $\gamma<1$, the situation is more complex, and without showing every step we exhibit the important results.
The relevant time scale for FRAP recovery is now
\begin{equation}
\tau_\gamma=\left(w^2/D_\gamma\right)^{1/\gamma}
\end{equation}
in place of Eq.\eqref{tau1}.  One has to solve the subdiffusion equation, the first equation in Eqs.~\eqref{evoleqc2} when $k_{on}^*$ and $k_{off}$ are set equal to zero. Proceeding through the Laplace transform process that led us to Eq.~\eqref{ns1}, we now arrive at the somewhat simpler form
\begin{equation}
\widetilde{frap}_{\gamma,PD}(p)
=\frac{2 F_{eq}}{p} K_1(p\tau_\gamma) I_1(p \tau_\gamma).
\label{frapLaplace}
\end{equation}

The result \eqref{frapLaplace} can be rewritten in terms of a Fox H-function \cite{Mathai1978}, which can then be analytically Laplace inverted to yield the FRAP recovery curve as a function of time.  The Fox H-function itself can be exhibited in terms of a variety of equivalent forms, one of which is
\begin{equation}
\widetilde{frap}_{\gamma,PD}(p)
 = \frac{z}{\sqrt{\pi}} H^{2,1}_{1,3}\left[p\tau_\gamma  \LM{(1/2-z,z)}{(1-z,z),(-z,z),(-1-z,z)}\right],
\end{equation}
where $z\equiv \gamma^{-1}$. This form can be inverted analytically to another Fox H-function that can again be written in a variety of ways \cite{Glckle1993}.  The simplest form yields the CTRW analog of the Soumpasis result for normal diffusion~\cite{Soumpasis1983}, namely,
\begin{equation}
{frap}_{\gamma,PD}(t)
= \frac{1}{\sqrt{\pi}} H^{2,1}_{2,3}\left[  \frac{w^2}{D_\gamma t^{\gamma}}  \LM{(1/2,1),(1,\gamma)}{(1,1),(0,1),(-1,1)}\right].
 \label{frapt3}
\end{equation}
For $\gamma=1$ this result reduces to Eq.~\eqref{gammais1}.  We note that this result for our circular initial condition should also be obtained from the subordination argument of Lubelski and Klafter~\cite{Lubelski2008} using their expression
\begin{equation}
frap_{\gamma,PD}(t)= \int_0^\infty dt' A(t',t)  frap_{1,PD}(D_\gamma t'/D_1),
\label{frapLK1}
\end{equation}
where $A(t',t)$ is the one-sided L\'evy function

\begin{equation}
A(t',t)=\frac{1}{t'}\sum_{n=0}^\infty \frac{(-1)^n}{\Gamma(1-\gamma-\gamma n)\Gamma(1+n)} \left(\frac{t'}{t^{1+\gamma}}\right)^{1+n}.
\end{equation}

The analytic solution \eqref{frapt3} is difficult to analyze in this general form.  It is, however, possible with considerable work to find analytic expressions for the long-time and short-time behaviors of this function. Here we only present the final results.  For long times we find
\begin{eqnarray}
&&frap_{\gamma,PD}(t) \approx  1-\frac{w^2}{4 \Gamma(1-\gamma)}\frac{\ln(D_\gamma t^\gamma/w^2)}{D_\gamma t^\gamma} \nonumber \\
&& +\frac{1}{8\Gamma(1-\gamma)}\left[2\gamma \,\psi(1-\gamma)+4\gamma_E-4\ln(2)-1\right]\frac{w^2}{K_\gamma t^{\gamma}} \nonumber \\
&&+{\cal O}\left(\frac{1}{t^{2\gamma}}\right)+{\cal O}\left(\frac{\ln(t^\gamma)}{t^{2\gamma}}\right),
\end{eqnarray}
where $\psi(\cdot)\equiv \Gamma'(\cdot)/\Gamma(\cdot)$ is the Digamma function, the logarithmic derivative of the Gamma function., and $\Gamma_E=0.57721\ldots$ is the Euler number. Note that when $\gamma\to 1$ the logarithmic term in $frap_{\gamma,PD}(t)$
vanishes since $\lim_{\gamma\to 1^-}\left[\Gamma(1-\gamma)\right]^{-1}=0$.  Using the fact that $\lim_{\gamma\to 1} \psi(1-\gamma)/\Gamma(1-\gamma)=-1$, we arrive at the result
\begin{equation}
\lim_{\gamma\to 1} frap_{\gamma,PD}
=1-\frac{\tau_1}{4t}
+ {\cal O}\left[\left(\frac{\tau_1}{t}\right)^2\right].
\end{equation}
This same result is also obtained when Eq.~\eqref{gammais1} is expanded for smalll $\tau_1/t$.  The limit $\gamma\to 1$ is therefore not singular.

For short times it is more convenient to return to Eq.~\eqref{frapLaplace} and expand it for large $p\tau_\gamma$. Term by term inversion then leads to the small-$t$ expansion
\begin{eqnarray}
frap_{\gamma PD}(t) &\sim & \frac{1}{\Gamma(\gamma/2+1)}\,\frac{\left(D_\gamma t^\gamma\right)^{1/2}}{w} \nonumber \\
&& -\frac{3}{8\Gamma(3\gamma/2+1)}\frac{\left(D_\gamma t^\gamma\right)^{3/2}}{w^3}+{\cal O}(t^{5\gamma/2}).
\end{eqnarray}
When $\gamma\to 1$ this yields the same result as does a direct expansion of Eq.~\eqref{gammais1}.

\section{Conclusions}
\label{sec6}
We have presented a model for FRAP recovery for proteins that might move subdiffusively and that bind and unbind in the cell nucleus.  This has required the construction of a reaction-subdiffusion equation that models not only motion, but also a reaction that describes both losses and gains (binding and unbinding). We have constructed this equation based on a continuous time random walk (CTRW) version of subdiffusion, and have suggested that this may be the first complete analytic model for FRAP recovery curves when the motion of the binding and unbinding entities is subdiffusive.

The motivation for this work began when we observed a contradiction in the literature.  On the one hand, the vast majority of models of motion of proteins and other entities in crowded environments such as a cell or a cell nucleus assume that the motion is subdiffusive.  There are a number of different models of subdiffusion in cells, and a great deal of discussion and even argument surrounds the question of which is the ``correct" model and even whether there is a single ``correct" model. However, on the other hand, the FRAP theory literature seems to be based entirely on diffusive motion, with very rare recognition or even mention of the subdiffusive motion paradigms. We wished to contribute toward filling this gap by generating a model for FRAP based on subdiffusive motion.

While we recognize that the issue of which model to use for subdiffusion is far from settled and the discussion rages on, for our purposes we have chosen a particular one of these models, namely, a CTRW, as noted above.  Not only have we done a great deal of work with CTRWs, but it is the only model that seems to make it possible to include reactions in the equation that describes the motion.
We pointed out that including reactions in a subdiffusion model is a complicated task because every microscopic situation leads to a different equation, and because in any case, subdiffusion and reactions are not simply additive as they are in a reaction-diffusion model.  The two components are intimately enmeshed. We derived the appropriate equation for the particular FRAP analysis that we wished to carry out.  We were then able to solve the problem analytically up to the time Laplace transform of the recovery curve.  From there, to obtain the time dependent curve required a numerical inversion.  This last step was the only one that required numerical work; all the other steps to this point are analytic.  In certain limiting cases we were able to carry out the inversion analytically.

Our purpose in carrying out this program was to compare our reaction-subdiffusion approach to a reaction-diffusion model in capturing the experimental results presented by Sprague et al.~\cite{Sprague2004}. The specific question we wished to address was whether a subdiffusion model is at least as good as a diffusion model for fitting FRAP data.  Both models have a number of parameters, subdiffusion one more than diffusion (the anomalous exponent $\gamma$), and optimizing the models with respect to all of them is a fairly extensive task.  Sprague et al. did this for the reaction-diffusion model.  We set some of our parameter values to be equal to those of the diffusion-based model and optimized with respect to only one or two.  In any case, the bottom line is that subdiffusion captures the experiments as well as does diffusion.  It is therefore appropriate to use a subdiffusion approach when working in a crowded environment where other measures have confirmed this slower motion.

A number of possible tasks remain to be carried out.  For instance, we can work with different geometries, different initial conditions, different inhomogeneities in the medium, and a number of other variations that have been considered in the diffusion-based literature~\cite{Sprague2004,Mueller2008,Hallen2010}. In Section~\ref{sec5} we conjectured that results for subdiffusion when the reaction is much faster than subdiffusion and when subdiffusion is very fast compared to both binding and to the time scale of the FRAP measurements would be similar to those found by Sprague et al.~\cite{Sprague2004} for normal diffusion.  These conjectures remain to be demonstrated. An extension of our one-binding-state model to an $n$-binding-state model when there are more than a single type of binding site is also possible. We continue to work on these and other extensions of this work.

\ack

This work was partially funded by the Ministerio de Ciencia y Tecnolog\'ia (Spain) through Grant No. FIS2010-16587 (partially financed by FEDER funds), by the Junta de Extremadura through Grant No. GRU10158, and by the US National Science Foundation under Grant No. PHY-0855471.

\section*{References}

     \bibliographystyle{unsrt}	


\begin{thebibliography}{10}

\bibitem{Sprague2004}
Sprague B L, Pego R L, Stavreva D A and McNally J G 2004
\newblock {Analysis of binding reactions by fluorescence recovery after  photobleaching}
\newblock{\em Biophys. J.}  \textbf{86} 3473--95

\bibitem{Dix2008}
 Dix J A and  Verkman A S 2008
\newblock {Crowding effects on diffusion in solutions and cells}
\newblock {\em Annu. Rev. Biophys.}  \textbf{37} 247--63

\bibitem{Sokolov2012}
 Sokolov I M 2012
\newblock {Models of anomalous diffusion in crowded environments}
\newblock {\em Soft Matter}  \textbf{8} 9043

\bibitem{Hofling2013}
  H\"{o}fling F and   Franosch T 2013
\newblock {Anomalous transport in the crowded world of biological cells}
\newblock {\em Rep. Prog. Phys.} \textbf{76} 046602

\bibitem{Boon2012}
 Boon J P,  Lutsko J and  Lutsko C 2012
\newblock {Microscopic approach to nonlinear reaction-diffusion: The case of  morphogen gradient formation}
\newblock {\em Phys. Rev. E} \textbf{85} 021126

\bibitem{Barkai2012}
 Barkai E,  Garini Y and  Metzler R 2012
\newblock {Strange kinetics of single molecules in living cells}
\newblock {\em Phys. Today} \textbf{65} 29

\bibitem{Yuste2008a}
Yuste S B  and Lindenberg K 2008
\newblock{\emph{Subdiffusion-limited A+A reactions}} in
 {\em Anomalous  Transport, Foundations and Applications} R. Klages, G. Radons and Sokolov I M (ed)  (Weinheim: Wiley-VCH)


\bibitem{Yuste2010a}
Yuste S B, Abad E and  Lindenberg K 2010
\newblock {Reaction-subdiffusion model of morphogen gradient formation}
\newblock {\em Phys. Rev. E} \textbf{82} 061123


\bibitem{Lubelski2008}
  Lubelski A and   Klafter J 2008
\newblock {Fluorescence recovery after photobleaching: the case of anomalous   diffusion}
\newblock {\em Biophys. J.} \textbf{94} 4646--53

\bibitem{Mathai1978}
 Mathai A M and Saxena  R K 1978 {\em {The H-Function with Applications in Statistics and other
  Disciplines}} (New York: Wiley)

\bibitem{Mendez2010}
 M\'{e}ndez V, Fedotov S, and Horsthemke W 2010
\newblock {\em {Reaction-Transport Systems: Mesoscopic Foundations, Fronts, and
  Spatial Instabilities}}
\newblock (Berlin:Springer-Verlag)


\bibitem{Vlad2002}
  Vlad M and   Ross J 2002
\newblock {Systematic derivation of reaction-diffusion equations with distributed delays and relations to fractional reaction-diffusion equations   and hyperbolic transport equations: Application to the theory of Neolithic   transition}
\newblock {\em Phys. Rev. E} \textbf{66} 061908

\bibitem{Seki2003}
  Seki K,   Wojcik M and Tachiya M 2003
\newblock {Fractional reaction-diffusion equation}
\newblock {\em J. Chem. Phys.} \textbf{119} 2165

\bibitem{Seki2003a}
  Seki K,   Wojcik M and Tachiya M 2003
\newblock {Recombination kinetics in subdiffusive media}
\newblock {\em J. Chem. Phys.} \textbf{119} 7525

\bibitem{Yadav2006}
Yadav A and  Horsthemke W 2006
\newblock {Kinetic equations for reaction-subdiffusion systems: Derivation and stability analysis}
\newblock {\em Phys. Rev. E} \textbf{74} 066118

\bibitem{Henry2006}
Henry B, Langlands T and Wearne S 2006
\newblock {Anomalous diffusion with linear reaction dynamics: From continuous time random walks to fractional reaction-diffusion equations}
\newblock {\em Phys. Rev. E} \textbf{74} 031116

\bibitem{Sokolov2006}
Sokolov I, Schmidt M and Sagu\'{e}s F 2006
\newblock {Reaction-subdiffusion equations}
\newblock {\em Phys. Rev. E} \textbf{73} 031102

\bibitem{Seki2007}
Seki K, Shushin A I,   Wojcik M and Tachiya M 2007
\newblock {Specific features of the kinetics of fractional-diffusion assisted geminate reactions}
\newblock {\em J. Phys. Condens. Matter} \textbf{\textbf{19}} 065117

\bibitem{Froemberg2008}
  Froemberg D and   Sokolov I 2008
\newblock {Stationary Fronts in an A+B$\to$0 Reaction under Subdiffusion}
\newblock {\em Phys. Rev. Lett.} \textbf{100} 108304

\bibitem{Henry2008}
Henry B, Langlands T and Wearne S 2008
\newblock {Fractional Cable Models for Spiny Neuronal Dendrites}
\newblock {\em Phys. Rev. Lett.} \textbf{100} 128103

\bibitem{Fedotov2010a}  Fedotov S  2010
Non-Markovian random walks and nonlinear reactions: Subdiffusion and propagating fronts
 \emph{Phys. Rev. E} \textbf{81} 011117


\bibitem{Yuste2012book}   Yuste S B, Abad E and Lindenberg K 2012
 \emph{Reactions in Subdiffusive Media and Associated Fractional Equations}
in  \emph{Fractional Dynamics: Recent Advances}  Klafter  J,
Lim  S C and  Metzler R (ed) (Singapore: World Scientific).

\bibitem{Soula2013}   Soula H, Car\'e B, Beslon G and Berry H 2013
Anomalous versus slowed-down Brownian diffusion in the ligand-binding equilibrium
\emph{Biophys. J.} \textbf{105} 2064--73


\bibitem{Shkilev2014}  Shkilev V P 2014
Comment on  ``Anomalous versus Slowed-Down Brownian Diffusion in the Ligand-Binding Equilibrium''
 \emph{ Biophys. J.} \textbf{106} 2541--3

\bibitem{Soula2014}  Soula H, Car\'e B, Beslon G and Berry H  2014
Reply to the Comment by V. P. Shkilev on ``Anomalous versus Slowed-Down Brownian Diffusion in the Ligand-Binding Equilibrium''
\emph{Biophys. J.} \textbf{106} 2544--6


\bibitem{Podlubny1999}   Podlubny  I 1999 \emph{Fractional Differential Equations: An Introduction to Fractional Derivatives, Fractional Differential Equations, to Methods of Their Solution and Some of Their Applications} (San Diego: Academic Press)

\bibitem{Kilbas2006}  Kilbas A A, Srivastava H M and Trujillo J J 2006 \emph{Theory And Applications of Fractional Differential Equations} (Amsterdam:Elsevier B.V.)


\bibitem{Mueller2008}
Mueller F, Wach P and McNally J G 2008
\newblock {Evidence for a common mode of transcription factor interaction with chromatin as revealed by improved quantitative fluorescence recovery after   photobleaching}
\newblock {\em Biophys. J.} \textbf{94}   3323--39




\bibitem{Carslaw1959}
  Carslaw  H  S  and  Jaeger  J   C 1959  \emph{Conduction of Heat in Solids} (Oxford: Oxford University Press)

\bibitem{Soumpasis1983} Soumpasis  D M 1983
Theoretical analysis of fluorescence photobleaching recovery experiments
\emph{Biophys. J.} \textbf{41} 95--7

\bibitem{Glckle1993}
 Gl\"{o}ckle W and  Nonnenmacher T F 1993
\newblock {Fox function representation of non-debye relaxation processes}
\newblock {\em J. Stat. Phys.} \textbf{71} 741--57

\bibitem{Hallen2010}
Hallen M A  and Layton  A T 2010
\newblock {Expanding the scope of quantitative FRAP analysis}
\newblock {\em J. Theor. Bio.} \textbf{262} 295--305


\end{thebibliography}


\end{document}